\documentclass[aps,prd,twocolumn,superscriptaddress,showpacs,floatfix]{revtex4}
\usepackage{bm}
\usepackage[dvips]{graphicx}
\usepackage{amsmath,amssymb,times}
\usepackage{ascmac}

\newcommand{\bequ}{\begin{equation}}
\newcommand{\eequ}{\end{equation}}
\newcommand{\bea}{\begin{eqnarray}}
\newcommand{\eea}{\end{eqnarray}}

\DeclareSymbolFont{boldletters}{OML}{cmm} {b}{it}
\DeclareSymbolFontAlphabet{\mathbit}{boldletters}
\DeclareMathSymbol{\alpha}{\mathalpha}{letters}{"0B}
\DeclareMathSymbol{\beta}{\mathalpha}{letters}{"0C}
\DeclareMathSymbol{\gamma}{\mathalpha}{letters}{"0D}
\DeclareMathSymbol{\delta}{\mathalpha}{letters}{"0E}
\DeclareMathSymbol{\epsilon}{\mathalpha}{letters}{"0F}
\DeclareMathSymbol{\zeta}{\mathalpha}{letters}{"10}
\DeclareMathSymbol{\eta}{\mathalpha}{letters}{"11}
\DeclareMathSymbol{\theta}{\mathalpha}{letters}{"12}
\DeclareMathSymbol{\iota}{\mathalpha}{letters}{"13}
\DeclareMathSymbol{\kappa}{\mathalpha}{letters}{"14}
\DeclareMathSymbol{\lambda}{\mathalpha}{letters}{"15}
\DeclareMathSymbol{\mu}{\mathalpha}{letters}{"16}
\DeclareMathSymbol{\nu}{\mathalpha}{letters}{"17}
\DeclareMathSymbol{\xi}{\mathalpha}{letters}{"18}
\DeclareMathSymbol{\pi}{\mathalpha}{letters}{"19}
\DeclareMathSymbol{\rho}{\mathalpha}{letters}{"1A}
\DeclareMathSymbol{\sigma}{\mathalpha}{letters}{"1B}
\DeclareMathSymbol{\tau}{\mathalpha}{letters}{"1C}
\DeclareMathSymbol{\upsilon}{\mathalpha}{letters}{"1D}
\DeclareMathSymbol{\phi}{\mathalpha}{letters}{"1E}
\DeclareMathSymbol{\chi}{\mathalpha}{letters}{"1F}
\DeclareMathSymbol{\psi}{\mathalpha}{letters}{"20}
\DeclareMathSymbol{\omega}{\mathalpha}{letters}{"21}
\DeclareMathSymbol{\varepsilon}{\mathalpha}{letters}{"22}
\DeclareMathSymbol{\vartheta}{\mathalpha}{letters}{"23}
\DeclareMathSymbol{\varpi}{\mathalpha}{letters}{"24}
\DeclareMathSymbol{\varrho}{\mathalpha}{letters}{"25}
\DeclareMathSymbol{\varsigma}{\mathalpha}{letters}{"26}
\DeclareMathSymbol{\varphi}{\mathalpha}{letters}{"27}
\DeclareMathSymbol{\Gamma}{\mathalpha}{letters}{"00}
\DeclareMathSymbol{\Delta}{\mathalpha}{letters}{"01}
\DeclareMathSymbol{\Theta}{\mathalpha}{letters}{"02}
\DeclareMathSymbol{\Lambda}{\mathalpha}{letters}{"03}
\DeclareMathSymbol{\Xi}{\mathalpha}{letters}{"04}
\DeclareMathSymbol{\Pi}{\mathalpha}{letters}{"05}
\DeclareMathSymbol{\Sigma}{\mathalpha}{letters}{"06}
\DeclareMathSymbol{\Upsilon}{\mathalpha}{letters}{"07}
\DeclareMathSymbol{\Phi}{\mathalpha}{letters}{"08}
\DeclareMathSymbol{\Psi}{\mathalpha}{letters}{"09}
\DeclareMathSymbol{\Omega}{\mathalpha}{letters}{"0A}

\begin{document}
\title{Quark number densities at imaginary chemical potential \\
in $N_f=2$ lattice QCD with Wilson fermions 
and its model analyses
}

\author{Junichi Takahashi}
\email[]{takahashi@phys.kyushu-u.ac.jp}
\affiliation{Department of Physics, Graduate School of Sciences, Kyushu University,
             Fukuoka 812-8581, Japan}

\author{Hiroaki Kouno}
\email[]{kounoh@cc.saga-u.ac.jp}
\affiliation{Department of Physics, Saga University,
             Saga 840-8502, Japan}

\author{Masanobu Yahiro}
\email[]{yahiro@phys.kyushu-u.ac.jp}
\affiliation{Department of Physics, Graduate School of Sciences, Kyushu University,
             Fukuoka 812-8581, Japan}

\date{\today}

\begin{abstract}
We investigate temperature ($T$) dependence 
of quark number densities ($n_{q}$) 
at imaginary and real chemical potential ($\mu$) by using $N_f=2$ 
lattice QCD and the hadron resonance gas (HRG) model. 
Quark number densities are calculated at imaginary $\mu$ with 
lattice QCD on an $8^{2}\times 16\times 4$ lattice 
with the clover-improved $N_f=2$ Wilson fermion action 
and the renormalization-group-improved Iwasaki gauge action. 
The results are consistent with the previous results of 
the staggered-type quark action.
The $n_{q}$ obtained are extrapolated to real $\mu$ by assuming 
the Fourier series for the confinement region and the polynomial series 
for the deconfinement region. 
The extrapolated results are consistent with the previous results of 
the Taylor expansion method for the reweighting factor. 
The upper bound $(\mu/T)_{\rm max}$ of the region where the extrapolation is 
considered to be reliable is estimated for each temperature $T$. 
We test whether $T$ dependence of nucleon and $\Delta$-resonance 
masses can be determined from LQCD data on $n_q$ at imaginary $\mu$ by using 
the HRG model. 
In the test calculation, nucleon and $\Delta$-resonance masses 
reduce by about 10\% in the vicinity of the pseudocritical temperature. 
\end{abstract}

\maketitle
\section{Introduction}
There are many interesting topics on quantum chromodynamics (QCD) at high density. 
The observation~\cite{Demorest} of a two-solar-mass neutron star 
has an impact on the equation of state (EoS) of dense matter and  
the QCD phase diagram at high density.
The experiments of relativistic heavy-ion collisions, for example 
the Beam Energy Scan experiments, are exploring  
QCD not only at finite temperature $T$ but also at finite quark-chemical 
potential $\mu$~\cite{STAR,PHENIX}.
Lattice QCD (LQCD) is the first-principle 
calculation to study QCD, but has the serious sign problem 
at finite $\mu$. 
\\
\indent 
In LQCD, the fermion determinant $\det M(\mu/T)$ becomes complex 
for finite real $\mu$, because 
\bea
(\det M(\mu/T))^{*}=\det M(-\mu^{*}/T)=\det M(-\mu/T). 
\label{sign-problem}
\eea
This interferes with the use of Monte-Carlo simulations 
based on the importance sampling. For this reason, several methods were 
proposed so far in order to avoid the sign problem~\cite{Muroya,Forcrand1}. 
Very recently, the complex Langevin method~\cite{Aarts1,Aarts2,Sexty,Aarts3} and the Lefschetz thimble theory~\cite{Cristoforetti,Fujii} have attracted much attention as the method to go beyond $\mu/T=1$.

\begin{figure}[h]
\begin{center}
\hspace{10pt}
 \includegraphics[angle=-90,width=0.445\textwidth]{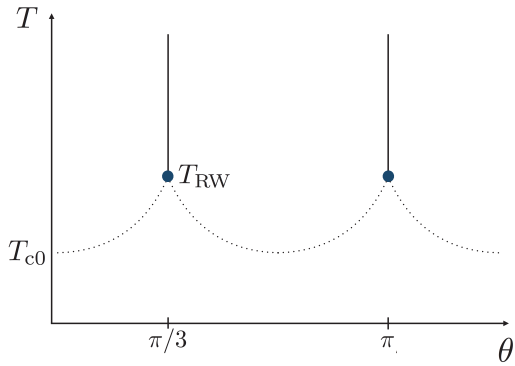}
\end{center}
\caption{
QCD phase diagram in the imaginary $\mu$ region. 
The solid and dashed lines stand for the RW phase transition and 
the deconfinement crossover, respectively. 
}
\label{QCDphase_muI}
\end{figure}

One of the methods to avoid the sign problem is 
the imaginary-$\mu$ approach. 
For pure imaginary chemical potential $\mu=i\mu_{\rm I}$, 
it is convenient to introduce the dimensionless chemical potential 
$\theta=\mu_{\mathrm{I}}/T$. The first equality of Eq. \eqref{sign-problem} 
shows that the fermion determinant $\det M(i\theta)$ is real 
for imaginary $\mu$. 
This makes LQCD simulations feasible there. 
Observables at real $\mu$ are extracted from those at imaginary $\mu$ 
by assuming functional forms for the observables. 
\\
\indent 
In the imaginary $\mu$ region, QCD has two characteristic properties: 
one is the Roberge-Weiss (RW) periodicity 
and the other is the RW phase transition~\cite{RW}. Figure~\ref{QCDphase_muI} shows a schematic graph for 
the QCD phase diagram in the $T$--$\theta$ plane. 
The QCD grand partition function $Z(\theta)$  has 
a periodicity of $2\pi/N_{c}$ in $\theta$: 
\bea
Z\left( \theta \right)=Z\left( \theta + \frac{2\pi k}{N_{c}} \right) ,
\eea
where $N_c$ is the number of colors and $k=1, \cdots, N_c$. 
This is a remnant of ${\mathbb Z}_{N_c}$ symmetry 
in pure gauge theory and is now called the RW periodicity. 
Meanwhile, the RW transition is the first-order phase transition appearing at 
$T$ higher than some temperature $T_{\rm RW}$ and $\theta=\pi/N_{c}$. 
This transition line and its ${\mathbb Z}_{N_c}$ images are plotted by 
the solid lines in Fig.~\ref{QCDphase_muI}. 
The point located at $(T,\theta)=(T_{\rm RW},\pi/N_{c})$ is 
called the RW endpoint.
Meanwhile, the dashed line represents the transition line 
of confinement/deconfinement crossover. 
The pseudocritical temperature $T_{\mathrm{c}}(\theta)$ 
is a function of $\theta$, and the value at $\theta=0$~\cite{Aoki} 
is denoted by $T_{\mathrm{c}0}$. 
As shown later, $T_{\mathrm{RW}}$ is located 
between $1.08T_{\mathrm{c}0}$ and $1.20T_{\mathrm{c}0}$, 
The order parameter of the RW transition is a ${\cal C}$-odd quantity 
such as the phase of the Polyakov loop or 
the quark-number density \cite{Kouno:2009bm}, 
where ${\cal C}$ means charge conjugation. 
The existence of the RW transition and the RW periodicity is numerically confirmed 
with LQCD simulations~\cite{Forcrand2,Wu,D'Elia1,D'Elia2,Forcrand3,Nagata} 
and the underlying mechanism is clearly understood 
with the effective model~\cite{Sakai:2008py,Kouno:2009bm,Sakai:2012} 
by introducing a new concept of extended ${\mathbb Z}_{N_c}$ symmetry. 
\\
\indent
The quark number density $n_{q}$ is a fundamental quantity 
to study high-density physics and important in determining the EoS at finite real $\mu$. 
The EoS plays an essential role in investigating the structure of neutron stars.
Moreover, $n_{q}$ is useful to determine the strength of 
vector-type interaction in the effective model~\cite{Sakai2008,Sugano}.
For small real $\mu/T$, the quark number density was calculated 
with the Taylor expansion method for the reweighting factor in which either 
the staggered-type~\cite{Allton} or the Wilson-type quark action~\cite{Ejiri} is taken. 
The quark number density is also computed at imaginary $\mu$ 
in Refs.~\cite{D'Elia1,D'Elia2,D'Elia3,Cea1,D'Elia4} with the staggered-type 
quark action, and $n_{q}$ at real $\mu$ is deduced from 
that at imaginary $\mu$ by assuming analytic forms for $n_{q}$. 
\\
\indent
In this paper, we investigate $\mu$ dependence of $n_{q}$ 
at both imaginary and real $\mu$. 
We first perform LQCD simulations at imaginary $\mu$ with the 
Wilson-type quark action, since the quark number density  at imaginary $\mu$ 
was not calculated with the Wilson-type quark action. 
LQCD simulations at imaginary $\mu$ do not require any special prescription 
in numerical calculations, 
since there is no sign problem. 
The $n_{q}$ obtained at imaginary $\mu$ are extrapolated to 
the real $\mu$ region by assuming functional forms for $n_{q}$. 
The extrapolated results are confirmed to be consistent with 
the previous results~\cite{Ejiri} of the Taylor expansion method 
for the reweighting factor. 
The upper bound $(\mu/T)_{\rm max}$ of the region where the extrapolation 
is considered to be reliable is estimated for each $T$. 
\\
\indent
The hadron resonance gas (HRG) model is reliable 
in the confinement region. 
For the 2+1 flavor case at zero chemical potential, in fact, 
it is shown in Ref.~\cite{Borsanyi:2012cr} that 
the model well reproduces LQCD data on pressure 
at $T<1.2T_{\mathrm{c}0}$. 
As shown by the dashed line in Fig. \ref{QCDphase_muI}, 
the pseudocritical temperature $T_{\mathrm{c}}(\theta)$ of 
deconfinement transition {\it increases} from $T_{\mathrm{c}0}$ to 
$T_{\rm RW}$ as $\theta$ increases from zero to $\pi/N_c$. 
As for real $\mu$, meanwhile, the pseudocritical temperature 
$T_{\mathrm{c}}(\mu)$ {\it decreases} as $\mu$ increases. 
When $\mu$ varies from pure imaginary value to real value 
with $T$ fixed at $T_{\mathrm{c}0}$, the system is thus 
in the confinement phase for imaginary $\mu$ and 
in the deconfinement phase for real $\mu$. 
Using this property, one can suggest that $T$ dependence of nucleon and 
$\Delta$-resonance masses in the vicinity of $T_{\mathrm{c}0}$ can be 
determined from $n_q$ at imaginary $\mu$ by using the HRG model. 
We test how the suggestion works in this paper. 
\\
\indent
Actual LQCD simulations are done on an $8^{2}\times 16 \times 4$ lattice 
with the clover-improved two-flavor Wilson fermion action and the renormalization-group-improved Iwasaki gauge action. 
We confirmed that the $n_q$ calculated 
on an $8^{2}\times 16 \times 4$ lattice are consistent with 
the previous results~\cite{D'Elia2,D'Elia4} calculated on a $16^3\times 4$ 
lattice. 
We then adopted an $8^{2}\times 16 \times 4$ lattice 
to reduce simulation time and take more trajectories. 
We consider two temperatures $T/T_{\mathrm{c}0}=$0.93 and 0.99 in the confinement 
region and four temperatures $T/T_{\mathrm{c}0}=$1.08, 1.20, 1.35 and 2.07 in the 
deconfinement region. 
Following the previous LQCD simulation~\cite{Ejiri}, we compute $n_{q}$ 
along the line of constant physics at $m_{\rm PS}/m_{\rm V}=0.80$, 
where $m_{\rm PS}$ and $m_{\rm V}$ are pseudoscalar- and vector-meson masses, 
respectively. 
This corresponds to the case of the pion mass $m_{\pi} \sim 616$~MeV and 
the quark mass $m_{q}\sim 130$~MeV~\cite{Sugano} for 
$T_{\mathrm{c}0}\sim 171$~MeV~\cite{Khan1}.
The analytic continuation is carried out with the Fourier series for 
$T<T_{\mathrm{c}0}$ and the polynomial series for $T>T_{\mathrm{RW}}$. 
\\
\indent 
This paper is organized as follows. 
In Sec.~\ref{Formulation}, we explain the lattice action, 
the quark number density and the analytic continuation.
In Sec.~\ref{Numerical results}, we show our simulation parameters and 
numerical results for $n_q$ at both imaginary and real $\mu$.  
In Sec.~\ref{HRG model}, we test whether $T$ dependence of nucleon and $\Delta$-resonance masses can be determined from LQCD data on $n_q$ at imaginary $\mu$ 
by using the HRG model.
Section~\ref{Summary} is devoted to a summary. 

\section{Formulation}
\label{Formulation}

\subsection{Lattice action}
We use the renormalization-group-improved Iwasaki gauge action $S_{g}$~\cite{Iwasaki} and the clover-improved two-flavor Wilson quark action $S_{q}$~\cite{clover-Wilson} defined by
\bea
S&=&S_{g}+S_{q}, \\
S_{g}&=&-\beta \sum_{x} \left(c_{0} \sum^{4}_{\mu<\nu;\mu,\nu=1}W^{1\times 1}_{\mu\nu}(x)\right. \nonumber \\
&&\left. +c_{1} \sum^{4}_{\mu\neq\nu;\mu,\nu=1}W^{1\times 2}_{\mu\nu}(x)\right), \\
S_{q}&=&\sum_{f=u,d}\sum_{x,y}\bar{\psi}^{f}_{x}M_{x,y}\psi^{f}_{y},
\eea
where $\beta=6/g^{2}$ for the gauge coupling $g$, 
$c_{1}=-0.331$, $c_{0}=1-8c_{1}$, and
\bea
&&M_{x,y}=
\nonumber \\
&&\delta_{xy}-\kappa\sum^{3}_{i=1}\{(1-\gamma_{i})U_{x,i}\delta_{x+\hat{i},y}+(1+\gamma_{i})U^{\dag}_{y,i}\delta_{x,y+\hat{i}}\} \nonumber \\
&&-\kappa\{e^{\hat{\mu}}(1-\gamma_{4})U_{x,4}\delta_{x+\hat{4},y}+e^{-\hat{\mu}}(1+\gamma_{4})U^{\dag}_{y,4}\delta_{x,y+\hat{4}}\} \nonumber \\
&&-\delta_{xy}c_{\mathrm{SW}}\kappa\sum_{\mu<\nu}\sigma_{\mu\nu}F_{\mu\nu}.
\eea
Here $\kappa$ is the hopping parameter, $\hat{\mu}$ is the quark chemical potential in lattice units, and the lattice field strength $F_{\mu\nu}$ is defined 
as $F_{\mu\nu}=(f_{\mu\nu}-f^{\dag}_{\mu\nu})/(8i)$ 
with $f_{\mu\nu}$ the standard clover-shaped combination of gauge links. 
For the clover coefficient $c_{\mathrm{SW}}$, we take the mean-field value 
estimated from $W^{1\times 1}$ in the one-loop level: 
$c_{\mathrm{SW}}=(W^{1\times 1})^{-3/4}=(1-0.8412\beta^{-1})^{-3/4}$~\cite{Iwasaki}. 
The value of $\kappa$ is determined at $\mu=0$ for each $\beta$ along the line of constant physics with $m_{\mathrm{PS}}/m_{\mathrm{V}} = 0.80$~\cite{Khan1,Khan2,Maezawa}.

\subsection{Quark number density}
The quark number density $n_{q}$ is defined as
\bea
\frac{n_{q}}{T^{3}} &=& \frac{1}{VT^{2}}\frac{\partial}{\partial \mu}\ln Z \label{eq:n_Z}\\
&=& \frac{N_{f}N_{t}^{3}}{N_{V}}\mathrm{tr}\left[M^{-1}\frac{\partial M}{\partial \hat{\mu}}\right], \label{eq:n}
\eea
where $V$ is the volume, $N_{f}$ is the number of flavors, $N_{t}$ is the temporal lattice size, $N_{V}$ is the lattice volume and $M$ is the fermion matrix.
We apply the random-noise method for the trace in Eq.~(\ref{eq:n}).
The number of noise vectors is about 4,000. 
The partition function $Z$ is $\mu$-even (${\cal C}$-even), so that 
$n_{q}$ is $\mu$-odd (${\cal C}$-odd) from Eq. \eqref{eq:n_Z}.
This means that $n_{q}$ is pure imaginary for imaginary $\mu$; 
actually, 
\bea
n_{q}^{\ast} = \left(\frac{1}{V}\frac{\partial \ln Z}
{\partial (i \theta)} \right)^{\ast} 
= \frac{1}{V}\frac{\partial \ln Z}{\partial (-i \theta)}
= -n_{q}.
\eea
We have confirmed in our LQCD simulations that the real part of $n_{q}$ is 
zero at imaginary $\mu$. For later convenience, we represent 
the imaginary part of $n_q$ by $n_q^{\rm I}$: 
$n_q^{\rm I}={\rm Im}(n_q)$.

\subsection{Analytic continuation}
Our final interest is $n_{q}$ at real $\mu$. 
We then extrapolate the $n_{q}$ calculated at imaginary $\mu$ with LQCD to the 
real $\mu$ region, assuming some functional forms for $n_{q}$. 
As for the pseudocritical line,  
it is shown in Refs.~\cite{Nagata,Cea2,Cea3,Cea4} that the terms 
of order higher than $\mu^{2}$ are necessary. 

In the imaginary-$\mu$ region, $n_{q}$ is a $\theta$-odd function 
with the RW periodicity. We then consider only a period 
$- \pi/3 < \theta \le \pi/3$ for simplicity. 
In the confinement region at $T<T_{\mathrm{c}0}$, the quark number density 
is smooth for any $\theta$, indicating that $n_{q}=0$ at 
$\theta=0, \pm \pi/3$~\cite{Sakai:2008py,Kouno:2009bm,Sakai:2012}. 
Hence $n_{q}$ can be described with good accuracy 
by a partial sum $S_{F}^{n}(T, \theta)$ of the Fourier series:
\bea
\frac{n_{q}(T,i \theta)}{T^{3}} \approx 
i S_{F}^{n}\left(T, \theta \right) =
i \sum_{k=1}^{n}a_{F}^{(k)}(T) \sin \left(3k
\theta \right) 
\label{eq:Fourier_sin}
\eea
where the superscript $n$ of $S_{F}^{n}(T,\theta)$ represents 
the highest order in the partial sum. 
The coefficients $a_{F}^{(k)}(T)$ are obtained by fitting the function 
\eqref{eq:Fourier_sin} to LQCD data at imaginary $\mu=i\mu_{\mathrm{I}}$. 
The analytic continuation from $\mu=i\mu_{\mathrm{I}}$ 
to $\mu=\mu_{\mathrm{R}}$ 
can be made by replacing $i \mu_{\mathrm{I}}/T$ 
by $\mu_{\mathrm{R}}/T$ in Eq. \eqref{eq:Fourier_sin}: 
\bea
\frac{n_{q}(T,\mu_{\mathrm{R}}/T)}{T^{3}} &\approx& 
g_{F}^{n}\left(T, \frac{\mu_{\mathrm{R}}}{T}\right)
\nonumber \\
&=&\sum_{k=1}^{n}a_{F}^{(k)}(T) \sinh \left(3k\frac{\mu_{\mathrm{R}}}{T}\right) .
\label{eq:Fourier_sinh}
\eea
Here note that the coefficients $a_{F}^{(k)}(T)$ have already been determined 
at imaginary $\mu$.

In the region $T_{\mathrm{c}0}<T<T_{\mathrm{RW}}$, the system is 
in the deconfinement region for small $\theta$ but in the confinement region 
for 
large $\theta$ near $\pi/3$, as shown in Fig. \ref{QCDphase_muI}. 
Because of this property, the $\theta$ dependence of $n_q$ is complicated and 
makes the analytic continuation difficult. We then do not perform 
the analytic continuation in this region.

In the deconfinement region at $T>T_{\mathrm{RW}}$, the quark number density 
is discontinuous at $\theta= \pm \pi/3$ where 
the first-order RW phase transition takes place; 
note that $n_q$ is the order parameter of the RW first-order 
transition~\cite{Kouno:2009bm}. 
Owing to this property, 
$n_q$ monotonically increases with $\theta$, as shown later 
in Fig. \ref{all-quark_number_density}.  
This suggests that $n_q$ can be described with good accuracy by 
a partial sum $S_{p}^{2n-1}(T, \theta)$ of the polynomial series: 
\bea
\frac{n_{q}(T,i \theta)}{T^{3}} \approx 
i S_{p}^{2n-1}\left(T, \theta \right) 
= i \sum_{k=1}^{n}a_{p}^{(2k-1)}(T) \theta^{2k-1} ,
\label{eq:poly}
\eea
where the superscript $n$ of $S_{p}^{2n-1}(T,\theta)$ represents 
the highest order in the partial sum. 
Again, the analytic continuation is made 
by replacing $i \mu_{\mathrm{I}}/T$ by $\mu_{\mathrm{R}}/T$ 
in Eq. \eqref{eq:poly}:
\bea
\frac{ n_{q}(T,\mu_{\mathrm{R}}/T) }{T^{3}}
&\approx& g_{p}^{2n-1}\left(T,\frac{\mu_{\mathrm{R}}}{T}\right)
\nonumber \\
&=&\sum_{k=1}^{n}(-1)^{(k-1)}a_{p}^{(2k-1)}(T)
\left(\frac{ \mu_{\mathrm{R} }}{T}\right)^{2k-1} .
\nonumber \\
\label{eq:poly-real}
\eea

\section{Numerical results}
\label{Numerical results}
Full QCD configurations with $N_f=2$ dynamical quarks were generated 
with the hybrid Monte-Carlo algorithm on a lattice of $N_{x}\times N_{y}\times N_{z}\times N_{t} = 8^2 \times 16\times 4$. 
The step size of the molecular dynamics is $\delta \tau = 0.02$ and 
the step number is $N_{\tau} = 50$.
The acceptance ratio is more than 95\%.
We generated about 32,000 trajectories and removed first 4,000 
trajectories for the thermalization of all the parameters, and measured $n_q$ at every 100 trajectories.
The relation of parameters $\kappa$ and $\beta$ to the corresponding $T/T_{\mathrm{c}0}$ was determined in Refs.~\cite{Khan1,Khan2,Maezawa}; see Table \ref{table-para} for the relation. 
\begin{table}[h]
\vspace{10pt}
\begin{center}
\vspace{5pt}
\begin{tabular}{|c|c|c|}
\hline 
$\kappa$ & $\beta$ & $T/T_{\mathrm{c}0}$
\\
\hline
~~$0.141139$~~ & ~~$1.80$~~ & ~~$0.93(5)$~~ \\
~~$0.140070$~~ & ~~$1.85$~~ & ~~$0.99(5)$~~ \\
~~$0.138817$~~ & ~~$1.90$~~ & ~~$1.08(5)$~~ \\
~~$0.137716$~~ & ~~$1.95$~~ & ~~$1.20(6)$~~ \\
~~$0.136931$~~ & ~~$2.00$~~ & ~~$1.35(7)$~~ \\
~~$0.135010$~~ & ~~$2.20$~~ & ~~$2.07(10)$~~ \\
\hline
\end{tabular}
\caption{
Summary of the simulation parameter sets determined in 
Refs.~\cite{Khan1,Khan2,Maezawa}. 
Note that $T_{\mathrm{c}0}\simeq 171$MeV, where $T_{\mathrm{c}0}$ is 
the pseudocritical temperature of deconfinement transition at $\mu=0$. 
In the parameter setting, 
the lattice spacing $a$ is about $0.14 \sim 0.2$~fm. }
\label{table-para}
\end{center}
\end{table}
\subsection{Quark number density at imaginary $\mu$}
\label{Sec:Quark number density at imaginary mu}
\begin{figure}[h]
\begin{center}
\vspace{10pt}
\hspace{10pt}
 \includegraphics[angle=-90,width=0.445\textwidth]{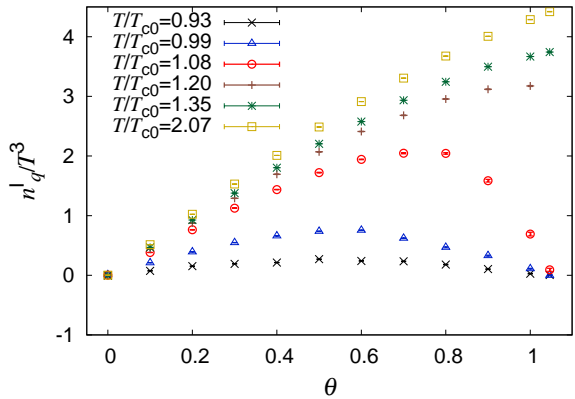}
\end{center}
\caption{
$\mu_{\rm I}/T$ dependence of $n^{\mathrm{I}}_{q}/T^{3}$ at various values of $T$.
The LQCD data are shown by symbols with error bars, 
although the error bars are quite small. 
}
\label{all-quark_number_density}
\end{figure}

Figure~\ref{all-quark_number_density} shows $n_{q}^{\rm I}/T^{3}$ 
as a function of $\theta$ for all the temperatures we consider. 
The LQCD data are plotted by symbols with error bars, 
although the errors are quite small. 
The number density $n_{q}^{\rm I}$ should be zero at $\theta=\pi/3$ 
below $T_{\rm RW}$ but finite above $T_{\rm RW}$, 
since $n_{q}^{\rm I}$ is the order parameter of the first-order RW phase 
transition. One can see from this fact 
that $T_{\rm RW}$ is located between $1.08T_{\mathrm{c}0}$ and 
$1.2T_{\mathrm{c}0}$. 
The quark number density $n_{q}^{\rm I}/T^{3}$ behaves as the sine function 
for $T< T_{\mathrm{c}0}$, but monotonically increases up to 
$\theta=\pi/3$ for $T > T_{\rm RW}$. 
As for $T=1.08T_{\mathrm{c}0}$, the system is in the deconfinement region 
for $\theta < 0.8$ but in the confinement region for $0.8 < \theta < \pi/3$, 
since $n_{q}^{\rm I}/T^{3}$ increases monotonically up to $\theta\sim 0.7$ but 
decreases to zero for $\theta>0.8$. This result is consistent with that of the 
staggered-type fermion in Ref.~\cite{D'Elia4}. The present result is thus 
independent of the fermion action taken.

First we consider the case of $T < T_{\mathrm{c}0}$ and 
determine the coefficients $a_{F}^{(k)}(T)$ of the Fourier series 
from the $n_q$ calculated at imaginary $\mu$ with LQCD. 
In principle, $n_q$ is described as an infinite series of sine functions 
for imaginary $\mu$ and of hyperbolic sine functions for real $\mu$, 
as shown in Eqs. \eqref{eq:Fourier_sin} and \eqref{eq:Fourier_sinh}. 
The partial sum is valid only when the series converges. 
Particularly for real $\mu$, the hyperbolic sine functions increase rapidly 
as $\mu/T$ becomes large. In this sense, it is important that 
the coefficients $a_{F}^{(k)}$ becomes small rapidly as $k$ increases.

\begin{figure}[h]
\begin{center}
\vspace{20pt}
\hspace{10pt}
 \includegraphics[angle=-90,width=0.45\textwidth]{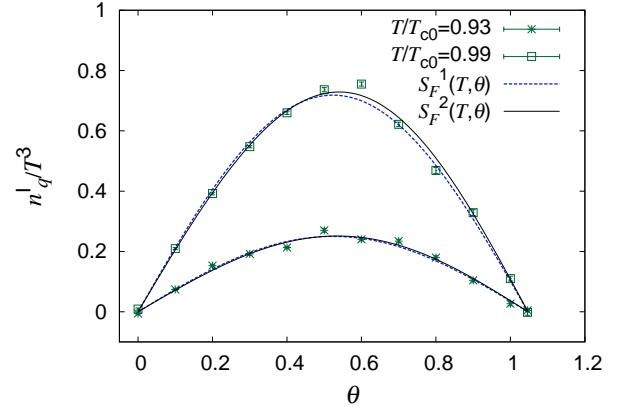}
\end{center}
\caption{
Results of $\chi^2$ fitting to LQCD data 
for the case of $T < T_{\mathrm{c}0}$. 
The results of $S_{F}^{1}$ and $S_{F}^{2}$ are plotted by 
dashed and solid lines, respectively. 
LQCD data are shown by symbols with error bars. 
}
\label{Fit_nmbdst-belowTc_mui_dep}
\end{figure}

In Fig.~\ref{Fit_nmbdst-belowTc_mui_dep}, 
the results of $\chi^2$ fitting are compared with the LQCD results. 
Here, two cases of $S_{F}^{1}$ and $S_{F}^{2}$ are plotted 
by dashed and solid curves, respectively. The two results well reproduce 
the LQCD data. 

The coefficients obtained are tabulated in Table~\ref{Fourier_coeff} 
for three cases of $S^{1}_{F}$, $S^{2}_{F}$ and $S^{3}_{F}$ and 
two cases of $T=0.93T_{\mathrm{c}0}$ and $0.99T_{\mathrm{c}0}$, 
together with the values of $\chi^{2}$ per degree of freedom (d.o.f.). 
As for $T=0.93T_{\mathrm{c}0}$, 
the absolute value of $a_{F}^{(2)}$ is much smaller than $a_{F}^{(1)}$ 
in $S_{F}^{2}$, but $a_{F}^{(3)}$ is comparable to the absolute value 
of $a_{F}^{(2)}$ in $S_{F}^{3}$. 
In addition, the $\chi^{2}/$d.o.f. value little changes 
between $S_{F}^{2}$ and $S_{F}^{3}$. 
The result of $S_{F}^{2}$ is thus acceptable, but that of $S_{F}^{3}$ is not. 
Similar discussion is possible for $T=0.99T_{\mathrm{c}0}$; 
note that the absolute value of $a_{F}^{(2)}$ is comparable
to that of $a_{F}^{(3)}$ in $S_{F}^{3}$  if the error ranges 
of $a_{F}^{(2)}$ and $a_{F}^{(3)}$ are taken into account. 
As shown in Fig.~\ref{Fit_nmbdst-belowTc_mui_dep}, 
moreover, the deviation of LQCD data from the solid line 
(the result of $S_{F}^{2}$) is rapidly oscillating 
with $\theta$ and cannot be reproduced by the next-order term $\sin(9\theta)$. 
Thus the coefficients higher than $a_{F}^{(2)}$ may not be determined from 
the present LQCD data. 
We then consider $S^{1}_{F}$ and $S^{2}_{F}$ 
as the extrapolation function from imaginary $\mu$ to real $\mu$. 

\begin{center}
\begin{figure}[h]
\vspace{10pt}
 \includegraphics[angle=-90,width=0.45\textwidth]{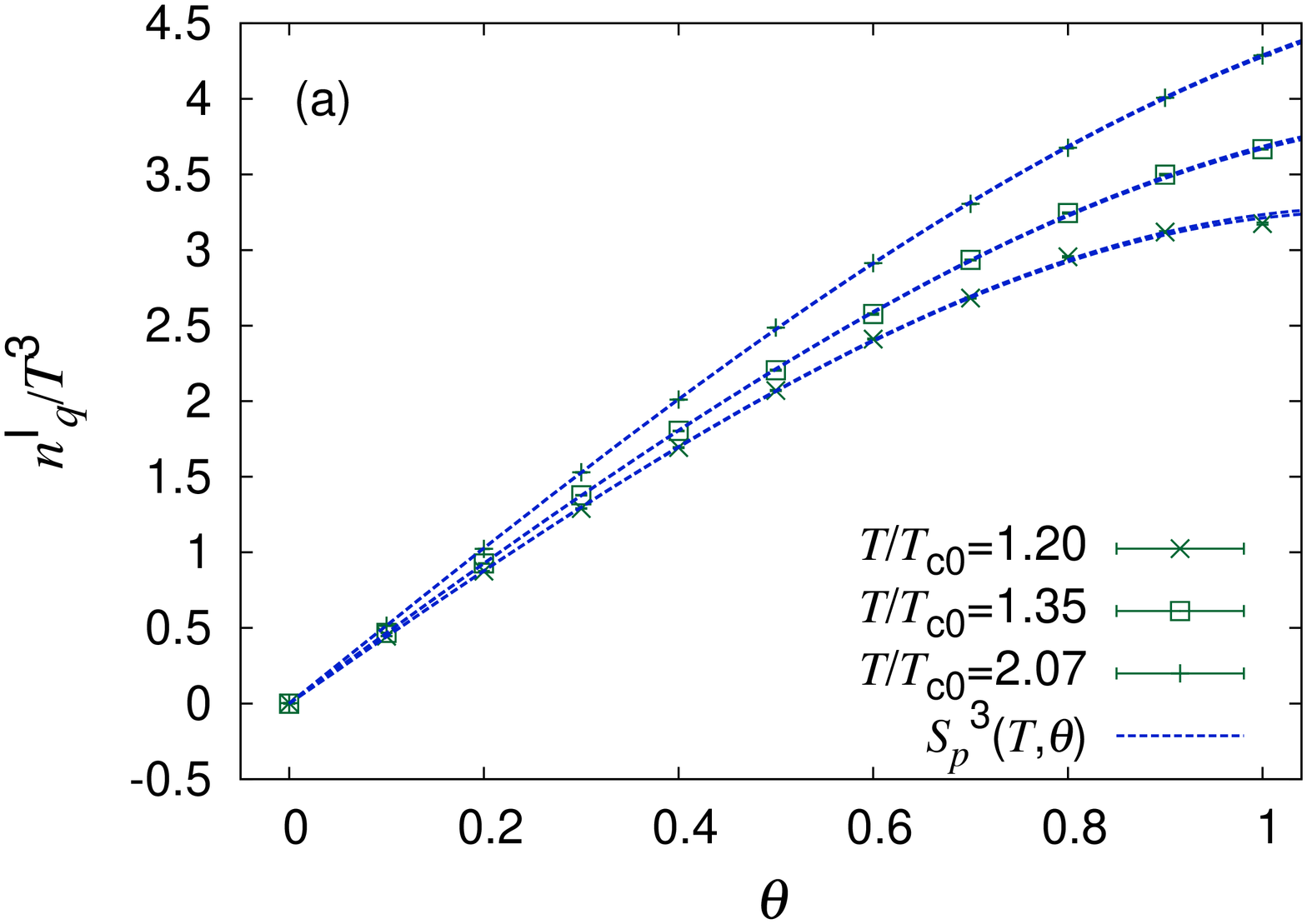}
\hspace{25pt}
 \includegraphics[angle=-90,width=0.45\textwidth]{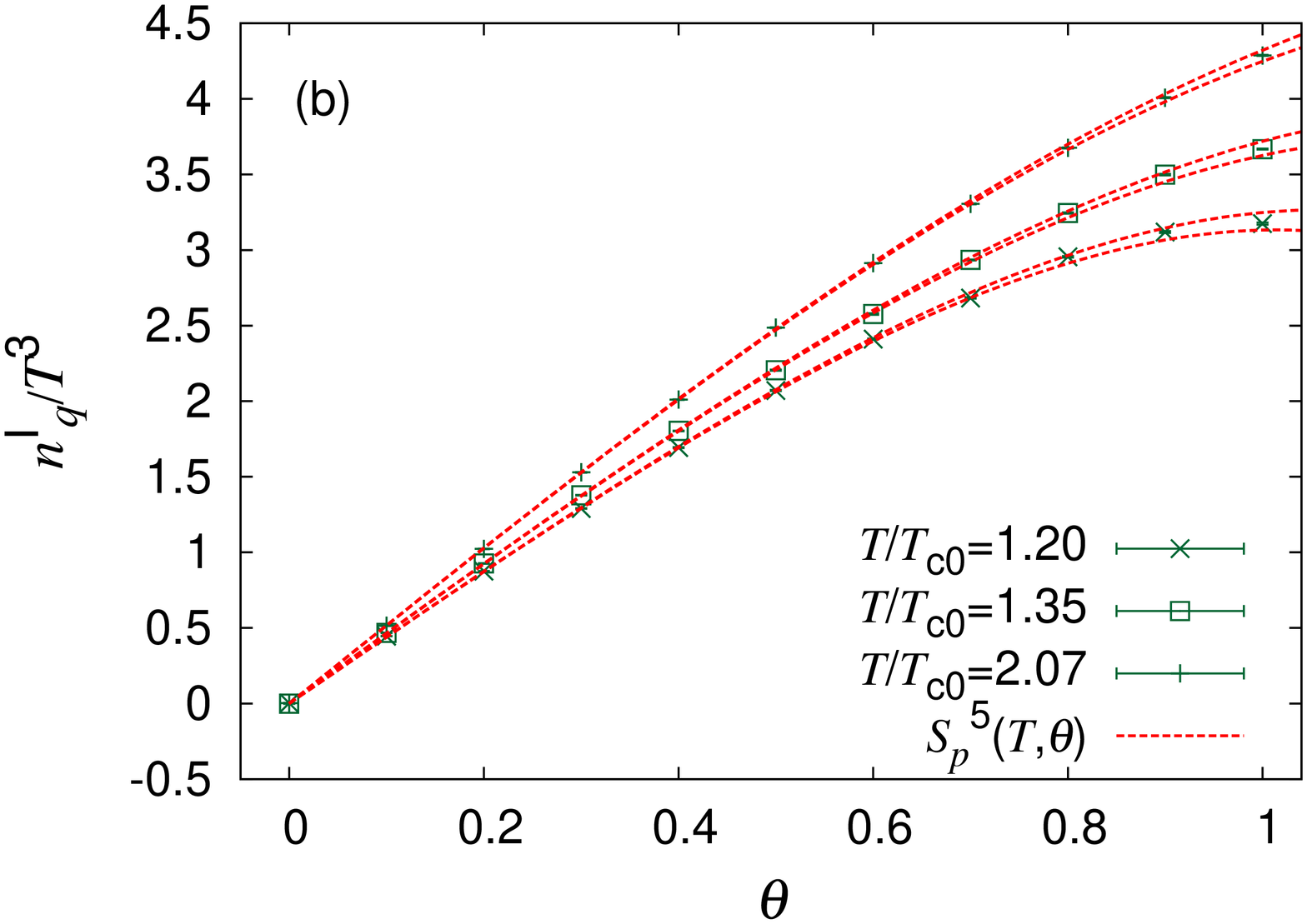}
\caption{
Results of $\chi^2$ fitting to LQCD data 
for the case of $T > T_{\mathrm{RW}}$. 
Panels (a) and (b) show the results of $S_{p}^{3}$ and $S_{p}^{5}$, 
respectively. 
For each temperature,  two dashed lines correspond to 
the upper and lower bounds of the $\chi^2$ fitting, respectively. 
LQCD data are shown by symbols with error bars. 
}

\label{Fit_nmbdst-aboveTc_mui_depn_p}
\end{figure}
\end{center}

Next we consider the case of $T > T_{\mathrm{RW}}$ and determine 
the coefficients $a_{p}^{(2k-1)}(T)$ of the polynomial series. 
In Fig.~\ref{Fit_nmbdst-aboveTc_mui_depn_p}, 
the fitting results are compared with LQCD data 
for $S^{3}_{p}$ in panel (a) and 
for $S^{5}_{p}$ in panel (b). 
For each case of $T=1.20T_{\mathrm{c}0}, 1.35T_{\mathrm{c}0}$ and 
$2.07T_{\mathrm{c}0}$, two dashed lines stand for 
the upper and lower bounds of $\chi^2$ fitting, respectively. 
The fitting results well reproduce LQCD data. 
The resulting coefficients $a_{p}^{(2k-1)}(T)$ are tabulated 
in Table~\ref{poly_coeff} for three cases of $S^{3}_{p}$, $S^{5}_{p}$ 
and $S^{7}_{p}$, together with the $\chi^{2}/$d.o.f. values. 
For each temperature, $S^{7}_{p}$ has the smallest $\chi^{2}/$d.o.f. value 
among $S^{3}_{p}$, $S^{5}_{p}$ and $S^{7}_{p}$. 
Particularly at $T=1.35T_{\mathrm{c}0}$, the value is almost one. 
Nevertheless, for each temperature the absolute value of $a_{p}^{(7)}$ is 
comparable to that of $a_{p}^{(5)}$ in $S_{p}^{7}$, whereas 
the absolute value of $a_{p}^{(5)}$ is smaller about by an order of 
magnitude than that of $a_{p}^{(3)}$ in $S_{p}^{5}$. 
We also performed a fit with ratios of polynomials to take account of 
the terms of order higher than $a_{p}^{(7)}$, following Ref. \cite{Cea3}. 
The resulting $\chi^{2}/$d.o.f. value is much larger than the case of 
$S_{p}^{7}$. 
This may indicate that $a_{p}^{(7)}$ and its higher-order coefficients 
cannot be determined properly from the present LQCD data. 
We then use $S_{p}^{3}$ and $S_{p}^{5}$ as the extrapolation function from 
imaginary $\mu$ to real $\mu$. 

\onecolumngrid
\begin{center}
\vspace{20pt}
\begin{table}[h]
\begin{tabular}{cccccc}
\hline 
$T/T_{\mathrm{c}0}$ & $a_{F}^{(1)}$ & $a_{F}^{(2)}$ & $a_{F}^{(3)}$ & $\chi^2/$d.o.f.~~ & ~~$\mu_{\mathrm{I}}/T$(fitting range)
\\
\hline
\hline
~~$0.93$~~&~~$0.250(2)$~&   ~               ~&~                ~&~$5.937$~&~$0\sim \pi/3$ \\
~~$0.93$~~&~~$0.251(2)$~&~~~~$-0.00457(216)$~&~                ~&~$6.084$~&~$0\sim \pi/3$ \\
~~$0.93$~~&~~$0.251(2)$~&~~~~$-0.00526(219)$~&~~~$0.00440(214)$~&~$6.290$~&~$0\sim \pi/3$ \\
\hline            
~~$0.99$~~&~~$0.718(2)$~&   ~               ~&~                ~&~$11.06$~&~$0\sim \pi/3$ \\
~~$0.99$~~&~~$0.728(3)$~&   ~$-0.0179(26)$  ~&~                ~&~$7.453$~&~$0\sim \pi/3$ \\
~~$0.99$~~&~~$0.727(3)$~&   ~$-0.0137(30)$  ~&~$-0.00825(276)$ ~&~$7.288$~&~$0\sim \pi/3$ \\
\hline            
\end{tabular}
\caption{
Coefficients of the Fourier series for $S^{1}_{F}$, $S^{2}_{F}$ and $S^{3}_{F}$. 
}
\label{Fourier_coeff}
\end{table}
\end{center}

\begin{center}
\vspace{25pt}
\begin{table}[h]
\begin{tabular}{cccccccc}
\hline 
$T/T_{\mathrm{c}0}$ & $a_{p}^{(1)}$ & $a_{p}^{(3)}$ & $a_{p}^{(5)}$ & $a_{p}^{(7)}$ & $\chi^2/$d.o.f.~~ & ~~$\mu_{\mathrm{I}}/T$(fitting range)
\\
\hline
\hline
$1.20$~&~$4.437(4)$~&      ~$-1.214(7)$   ~&        ~                ~&   ~               ~& $13.66$ & $0\sim 1$  \\
$1.20$~&~$4.407(5)$~&    ~~~$-1.024(27)$  ~&        ~$-0.1935(260)$  ~&   ~               ~& $8.472$ & $0\sim 1$  \\
$1.20$~&~$4.427(7)$~&    ~~~$-1.274(66)$  ~&      ~~~$-0.4458(1569)$ ~&  ~~$-0.4229(1024)$~& $7.245$ & $0\sim 1$  \\
\hline
$1.35$~&~$4.675(3)$~& ~~~~~~$-0.9973(49)$ ~&        ~                ~&   ~               ~& $6.036$ & $0\sim 1$  \\
$1.35$~&~$4.662(5)$~&~~~~~~~$-0.9223(223)$~&     ~~~~$-0.06736(1956)$~&   ~               ~& $5.308$ & $0\sim 1$  \\
$1.35$~&~$4.695(7)$~&   ~~~~$-1.295(67)$  ~&      ~~~$-0.7986(1469)$ ~&   ~$-0.5310(893)$ ~& $1.011$ & $0\sim 1$  \\
\hline            
$2.07$~&~$5.174(2)$~& ~~~~~~$-0.8904(40)$ ~&        ~                ~&   ~               ~& $9.161$ & $0\sim 1$  \\
$2.07$~&~$5.177(4)$~&~~~~~~~$-0.9056(177)$~&~~~~~~~~~$0.01356(1531)$ ~&   ~               ~& $10.21$ & $0\sim 1$  \\
$2.07$~&~$5.158(6)$~&~~~~~~~$-0.7119(432)$~&    ~~~~~$0.4381(932)$   ~&~~~~$0.2819(574)$  ~& $8.220$ & $0\sim 1$  \\
\hline            
\end{tabular}
\caption{
Coefficients of the polynomial series for $S^{3}_{p}$, $S^{5}_{p}$ and $S^{7}_{p}$. 
}
\label{poly_coeff}
\end{table}
\end{center}

\twocolumngrid

\subsection{Quark number density at real $\mu$}
\label{Quark number density at real mu} 
\begin{figure}[h]
\begin{center}
\vspace{10pt}
 \includegraphics[angle=-90,width=0.445\textwidth]{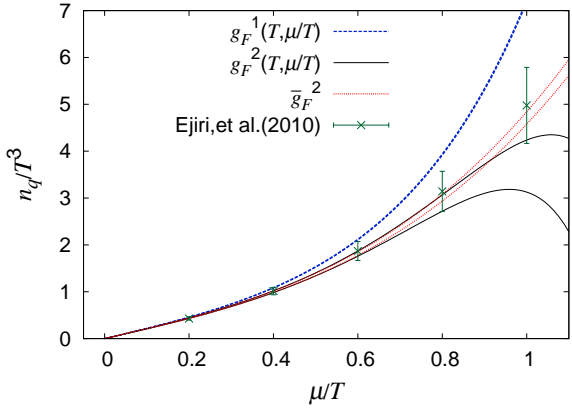}
\end{center}
\caption{
$\mu/T$ dependence of $n_{q}/T^{3}$ at $T=0.99T_{\mathrm{c}0}$. 
The results of $g_{F}^{1}$, $g_{F}^{2}$ and $\bar{g}_{F}^{2}$ are plotted 
by dashed, solid and dotted lines, respectively; 
see the text for the definition of $\bar{g}_{F}^{2}$. 
For each case, the upper and the lower bounds of $\chi^{2}$ fittings are 
shown by a pair of lines. 
The symbols denotes LQCD results of the Taylor expansion method 
for the reweighting factor in Ref.~\cite{Ejiri}. 
}
\label{Fit_nmbdst-b185mur_dep}
\end{figure}

First we consider the case of $T/T_{\mathrm{c}0}<1$. 
As the extrapolation function from imaginary $\mu$ to real $\mu$, 
we consider $g_{F}^{1}$ and $g_{F}^{2}$. 
Figure~\ref{Fit_nmbdst-b185mur_dep} shows $\mu/T$ dependence 
of $n_{q}/T^{3}$ for $T=0.99T_{\mathrm{c}0}$. 
The result of the extrapolation is shown by a pair of lines; 
the two lines correspond to the upper and lower bounds of the extrapolation 
and the uncertainty comes from the errors in $a_{F}^{(k)}$. 
The $g_{F}^{2}$ case (solid line) has a larger error 
than the $g_{F}^{1}$ case (dashed line), 
because the former error comes from both of $a_{F}^{(1)}$ and $a_{F}^{(2)}$ 
but the latter does from $a_{F}^{(1)}$ only. 
We also show the previous LQCD result~\cite{Ejiri} of 
the Taylor expansion method for the reweighting factor by symbols 
with error bars. In the previous calculation, $n_{q}$ is described 
by a polynomial series of $\mu/T$ and 
the terms up to $(\mu/T)^3$ are taken into account. 
The result of $g_{F}^{2}$ deviates from that of the Taylor expansion method 
at $\mu/T > 0.8$. 
To clarify what causes the deviation, in Eq. \eqref{eq:Fourier_sinh} for 
$g_{F}^{2}$ we expand 
the hyperbolic sine function into a polynomial series 
and discard the terms of order higher than $(\mu/T)^3$. 
We denote the resulting function by $\bar{g}_{F}^{2}$. 
The result of $\bar{g}_{F}^{2}$ (dotted line) 
is consistent with that of the Taylor expansion method at 
$\mu/T < 1$. 
Thus the difference between $g_{F}^{2}$ and 
the Taylor expansion method comes from the terms of order higher than 
$(\mu/T)^3$, and $g_{F}^{2}$ yields a correction to the result of 
the Taylor expansion. 
From the fact that the correction is small at $\mu/T < 0.8$, we can conclude 
that both the previous result of the Taylor expansion method and the present 
result of $g^{2}_{F}$ are reliable at least at $\mu/T < 0.8$.

\begin{figure}[h]
\begin{center}
\vspace{10pt}
 \includegraphics[angle=-90,width=0.445\textwidth]{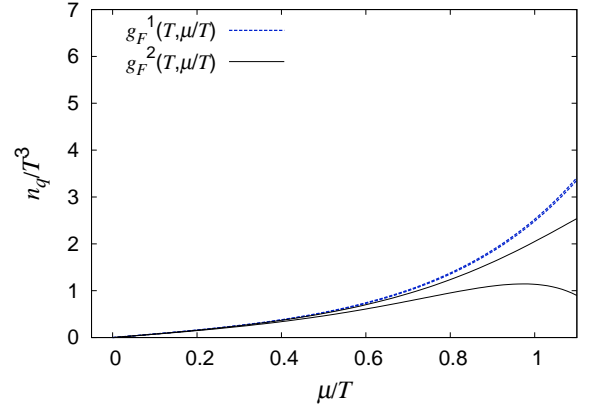}
\end{center}
\caption{
$\mu/T$ dependence of $n_{q}/T^{3}$ at $T=0.93T_{\mathrm{c}0}$. 
See Fig. \ref{Fit_nmbdst-b185mur_dep} for the definition of lines. 
Two cases of $g_{F}^{1}$ and $g_{F}^{2}$  are plotted. 
}
\label{Fit_nmbdst-b180mur_dep}
\end{figure}

Figure~\ref{Fit_nmbdst-b180mur_dep} shows $\mu/T$ dependence 
of $n_{q}/T^{3}$ at $T=0.93T_{\mathrm{c}0}$. 
The definition of lines is the same as 
in Fig. \ref{Fit_nmbdst-b185mur_dep}. 
The difference between two results of $g^{1}_{F}$ and 
$g^{2}_{F}$ reduces as $T$ decreases from $0.99T_{\mathrm{c}0}$ to $0.93T_{\mathrm{c}0}$. 
The extrapolation of $g^{2}_{F}$ thus becomes more reliable 
as $T$ decreases. 
As mentioned above, the $g^{2}_{F}$ extrapolation is 
reliable at least at $\mu/T < 0.8$ for $T=0.99T_{\mathrm{c}0}$. 
This means that also for $T=0.93T_{\mathrm{c}0}$ 
the $g^{2}_{F}$ extrapolation is reliable at least at $\mu/T < 0.8$. 

\begin{figure}[h]
\begin{center}
\vspace{10pt}
 \includegraphics[angle=-90,width=0.445\textwidth]{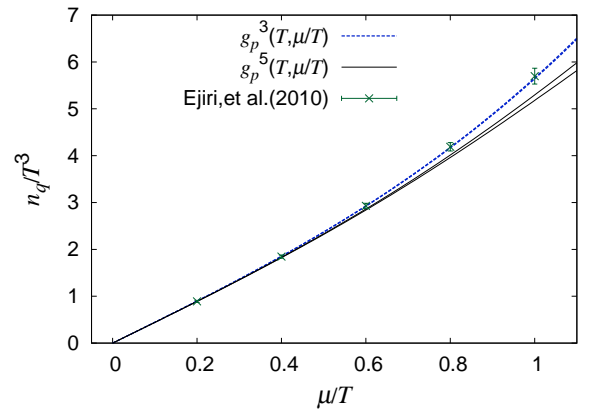}
\end{center}
\caption{
$\mu/T$ dependence of $n_{q}/T^{3}$ at $T=1.20T_{\mathrm{c}0}$.
Two cases of $g_{p}^{3}$ and $g_{p}^{5}$ are plotted. 
The upper and the lower bounds of $\chi^{2}$ fitting 
are shown by a pair of dashed lines for $g_{p}^{3}$ 
and by a pair of solid lines for $g_{p}^{5}$.
The symbols denotes LQCD results of the Taylor expansion method 
for the reweighting factor in Ref.~\cite{Ejiri}. 
}
\label{Fit_nmbdst-b195mur_dep}
\end{figure}

\begin{figure}[h]
\begin{center}
\vspace{10pt}
 \includegraphics[angle=-90,width=0.445\textwidth]{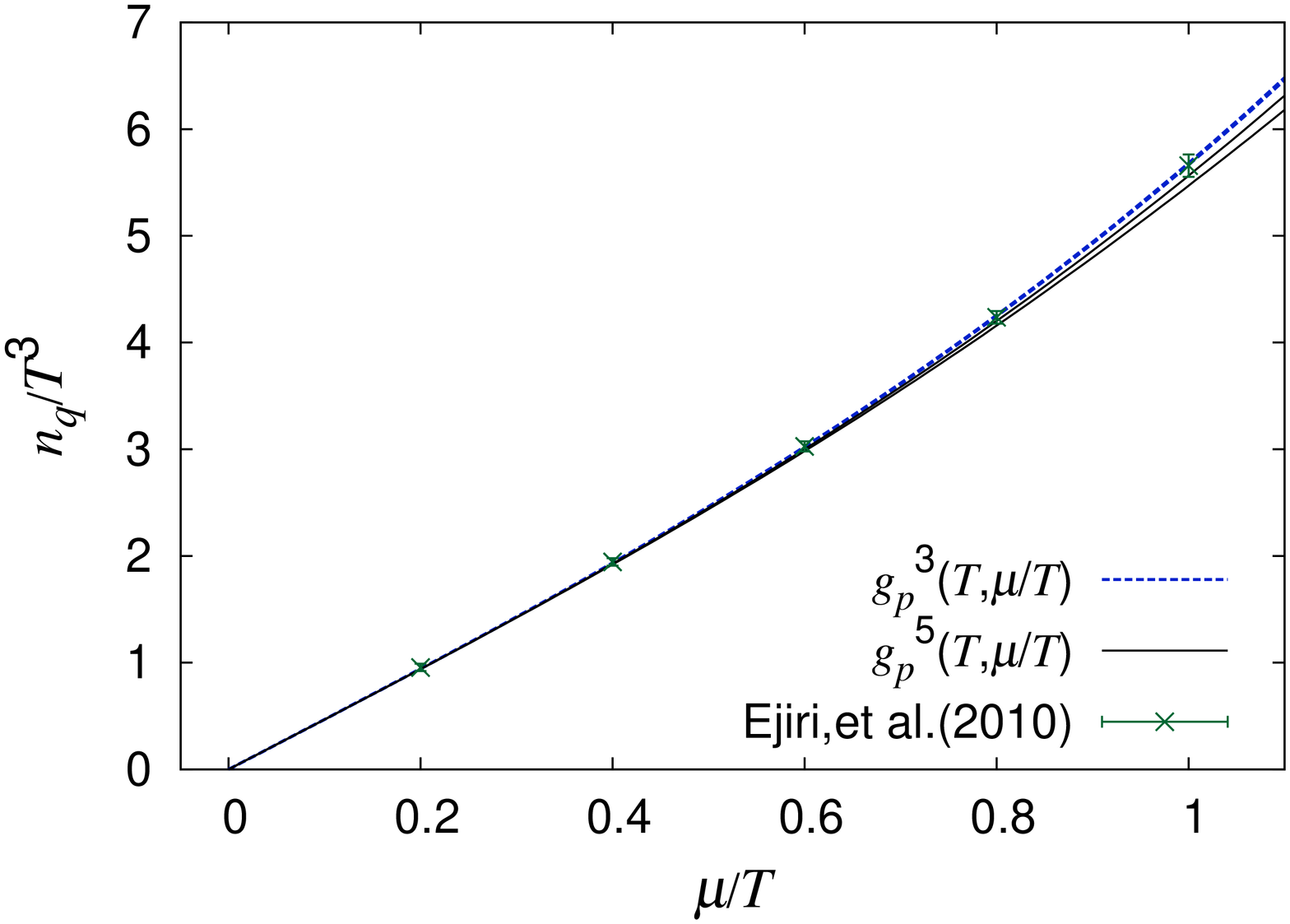}
\end{center}
\caption{
$\mu/T$ dependence of $n_{q}/T^{3}$ at $T=1.35T_{\mathrm{c}0}$. 
Two cases of $g_{p}^{3}$ and $g_{p}^{5}$ are plotted. 
See Fig. \ref{Fit_nmbdst-b195mur_dep} for the definition of lines. 
The symbols denotes LQCD results of the Taylor expansion method 
for the reweighting factor in Ref.~\cite{Ejiri}. 
}
\label{Fit_nmbdst-b200mur_dep}
\end{figure}

\begin{figure}[h]
\begin{center}
\vspace{10pt}
 \includegraphics[angle=-90,width=0.445\textwidth]{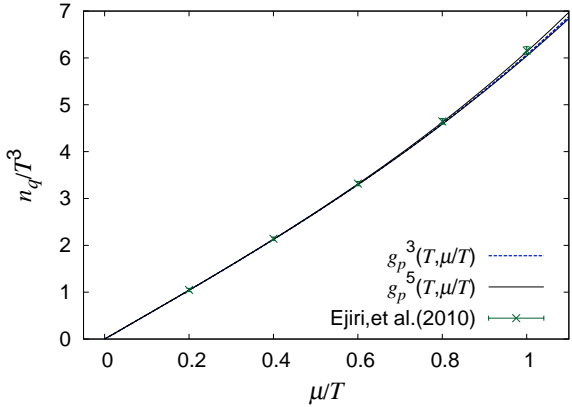}
\end{center}
\caption{
$\mu/T$ dependence of $n_{q}/T^{3}$ at $T=2.07T_{\mathrm{c}0}$.
Two cases of $g_{p}^{3}$ and $g_{p}^{5}$ are plotted. 
See Fig. \ref{Fit_nmbdst-b195mur_dep} for the definition of lines. 
The symbols denotes LQCD results of the Taylor expansion method 
for the  reweighting factor in Ref.~\cite{Ejiri}. 
}
\label{Fit_nmbdst-b220mur_dep}
\end{figure}

Next we consider the case of $T > T_{\rm RW}$. 
As the extrapolation function from imaginary $\mu$ to real $\mu$, 
we consider $g^{3}_{p}$ and $g^{5}_{p}$. 
Figures~\ref{Fit_nmbdst-b195mur_dep}, \ref{Fit_nmbdst-b200mur_dep} and \ref{Fit_nmbdst-b220mur_dep} show $\mu/T$ dependence of $n_{q}/T^{3}$ 
at $T=1.20T_{\mathrm{c}0}$, $1.35T_{\mathrm{c}0}$ and $2.07T_{\mathrm{c}0}$, 
respectively.
The result of $g^{3}_{p}$ ($g^{5}_{p}$) is plotted by a pair of dashed 
(solid) lines; the two lines correspond to the upper and lower bounds of 
extrapolation. 
For each temperature, the result of $g^{3}_{p}$ 
agrees with the previous result~\cite{Ejiri} 
of the Taylor expansion method for the reweighting factor.  
In the Taylor expansion method, $n_q$ is described by a polynomial series and 
the terms up to $(\mu/T)^3$ are taken. The highest order taken is 
the same between the $g^{3}_{p}$ extrapolation and the Taylor expansion 
method. The agreement of $n_{q}$ between the two methods 
is thus natural, although 
the coefficients are determined 
with different procedures between the two methods. 
The difference between $g^{3}_{p}$ and $g^{5}_{p}$ becomes small 
as $T$ increases. Contributions of order higher than $(\mu/T)^3$ thus become 
small as $T$ increases. 

\begin{center}
\begin{figure}[h]
\vspace{10pt}
 \includegraphics[angle=-90,width=0.45\textwidth]{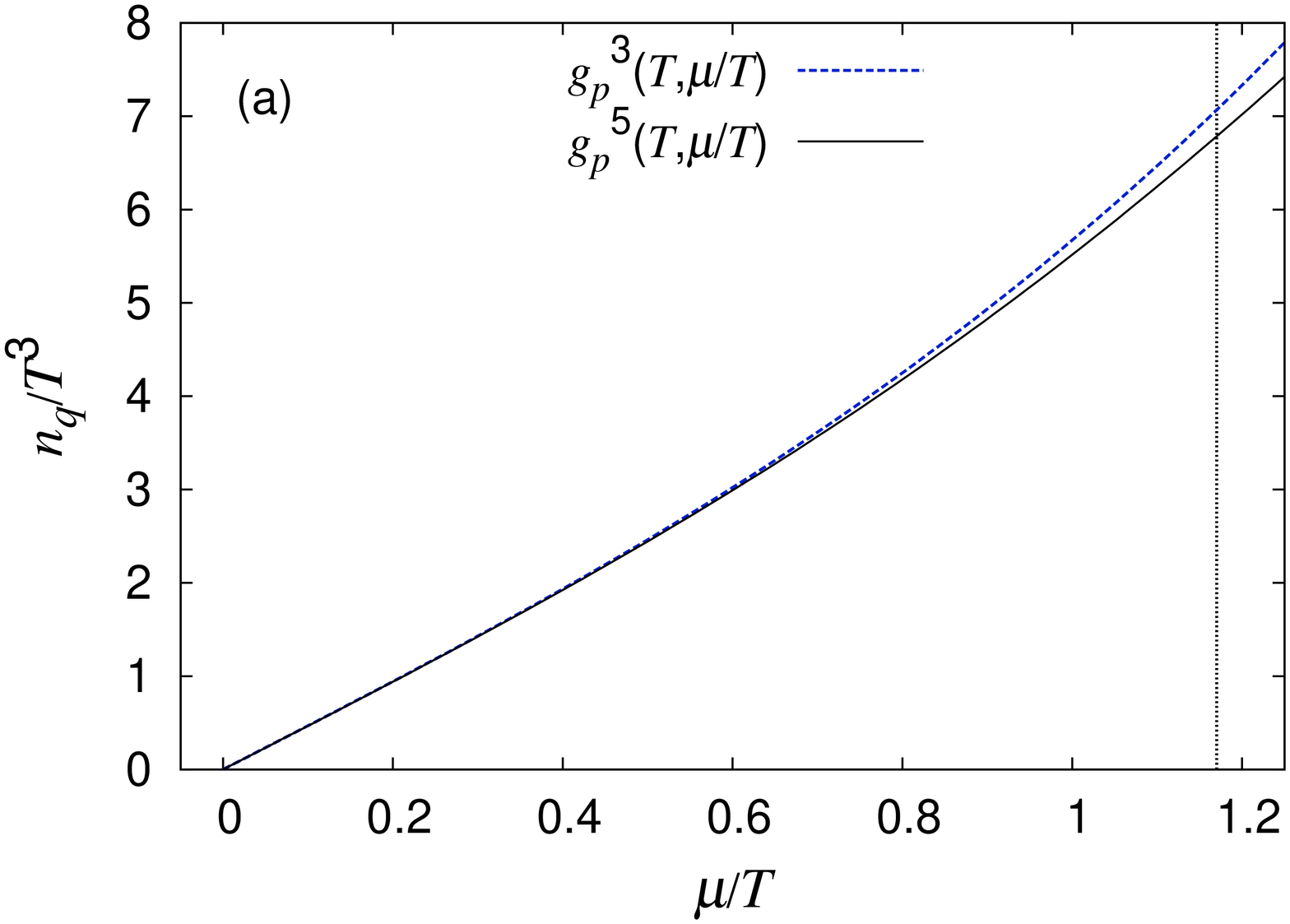}
\hspace{25pt}
 \includegraphics[angle=-90,width=0.45\textwidth]{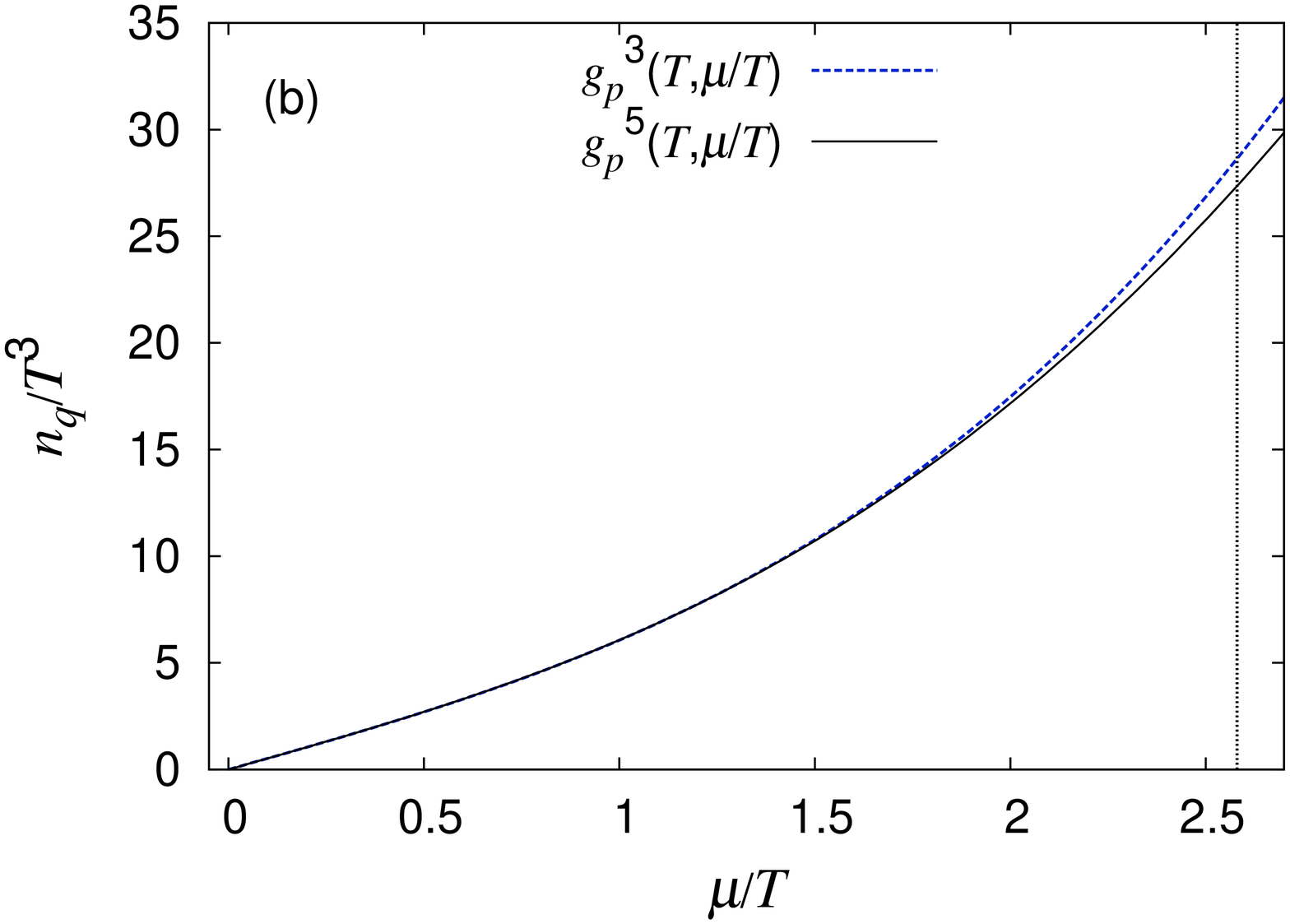}
\caption{
$\mu/T$ dependence of $n_{q}/T^{3}$ and 
the region of $\delta \le 0.1$ for 
(a) $T=1.35T_{\mathrm{c}0}$ and (b) $T=2.07T_{\mathrm{c}0}$. 
The relative difference $\delta$ exceeds $10\%$ 
at $\mu/T=(\mu/T)_{\mathrm{max}}$ represented by the vertical dotted line. 
The dashed and solid lines denote the mean values of $g_{p}^{3}$ and 
$g_{p}^{5}$, respectively.
}
\label{max_nmbdst}
\end{figure}
\end{center}

\begin{figure}[h]
\begin{center}
\vspace{10pt}
 \includegraphics[angle=-90,width=0.445\textwidth]{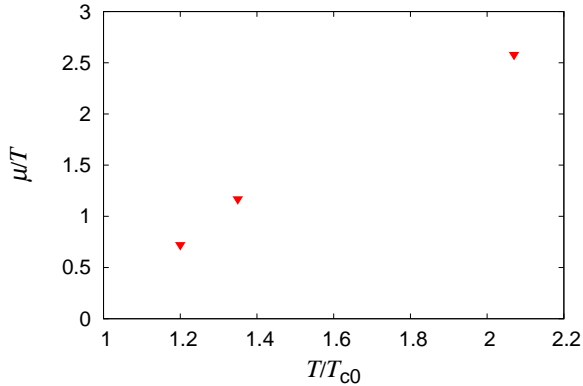}
\end{center}
\caption{
$T$ dependence of the upper bound $(\mu/T)_{\rm max}$ 
of the reliable extrapolation for the case of $T > T_{\rm RW}$. 
}
\label{upper-bound-of-region}
\end{figure}

Now we estimate the upper bound $(\mu/T)_{\rm max}$
of the region in which the extrapolation 
is considered to be reliable, and investigate $T$ dependence 
of $(\mu/T)_{\rm max}$ for $T > T_{\mathrm{RW}}$. 
For this purpose, we define the relative difference $\delta$ 
between $g^{3}_{p}$ and $g^{5}_{p}$ 
as $\delta = |g^{5}_{p}-g^{3}_{p}|/g^{5}_{p}$, and assume 
that the extrapolation is reliable when $\delta < 0.1$.
The relative difference $\delta$ exceeds $10\%$ 
at $\mu/T \simeq 0.72$ for $T=1.20T_{\mathrm{c}0}$, 
$\mu/T \simeq 1.2$ for $T=1.35T_{\mathrm{c}0}$ 
and $\mu/T \simeq 2.6$ for $T=2.07T_{\mathrm{c}0}$, 
as shown in Fig.~\ref{max_nmbdst}.
The upper bound $(\mu/T)_{\rm max}$ of the reliable extrapolation 
is plotted as a function of $T/T_{\mathrm{c}0}$ in Fig. \ref{upper-bound-of-region}. 
The upper bound goes up as $T$ increases, indicating that 
contributions of order higher than $(\mu/T)^3$ become less important 
as $T$ becomes high. 
Thus the present result of $g^{3}_{p}$ and the previous result~\cite{Ejiri} 
of the Taylor expansion method become more reliable as $T$ increases.

\section{Hadron resonance gas model}
\label{HRG model}
\begin{figure}[h]
\begin{center}
\vspace{10pt}
\hspace{10pt}
 \includegraphics[angle=-90,width=0.445\textwidth]{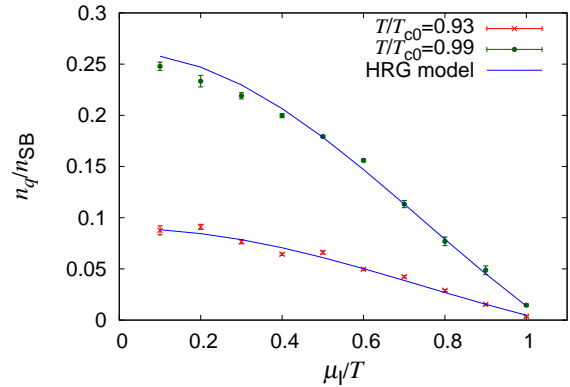}
\end{center}
\caption{
$\theta$ dependence of $n_{q}/n_{\mathrm{SB}}$ 
at $T=0.93T_{\mathrm{c}0}$ and $0.99T_{\mathrm{c}0}$. 
The cross and square symbols with error bars represent the LQCD results 
at $T=0.93T_{\mathrm{c}0}$ and $0.99T_{\mathrm{c}0}$, respectively, whereas  
the solid lines stand for the HRG model results. 
}
\label{HRG_im-nmb}
\end{figure}

Now we consider the confinement region at $\theta \ge 0$, and 
test whether nucleon and $\Delta$-resonance masses can be determined 
from the present LQCD results at $\theta \ge 0$ 
by using the HRG model particularly 
in the vicinity of $T_{\mathrm{c}0}$. 
The HRG model considers non-interacting hadrons and resonances, 
each classified with species $i$, i.e., 
with mass $m_{i}$, baryon number $B_{i}$ and isospin $I_{3i}$. 
For the 2+1 flavor case at zero chemical potential, the HRG model well reproduces LQCD data on 
pressure at $T<1.2T_{\mathrm{c}0}$~\cite{Borsanyi:2012cr}. 
This means that the HRG model is applicable at $T<T_{\mathrm{c}0}$ even if 
$\theta$ is finite, because $T_{\rm c}(\theta) > T_{\mathrm{c}0}$ 
for any finite $\theta$. The pressure of the model is obtained by 
\bea
p^{\mathrm{HRG}}&=&-\frac{T}{V}\sum_{i\in \mathrm{meson}}\ln Z^{{\rm M}}_{i}(T,V,\mu_{i}) \nonumber \\
&& -\frac{T}{V}\sum_{i\in \mathrm{baryon}}\ln Z^{{\rm B}}_{i}(T,V,\mu_{i}) 
\eea
with 
\bea
\ln Z^{{\rm M}/{\rm B}}_{i}=\pm \frac{Vg_{i}}{2\pi^{2}}\int^{\infty}_{0}dp p^{2} \ln\left(1\mp z_{i}e^{-\epsilon_{i}/T}\right)  
\eea
for the energy $\epsilon_{i}=\sqrt{p^{2}+m_{i}}$, the degeneracy factor $g_{i}$ and the fugacity
\bea
z_{i}=e^{\mu_{i}/T}=\exp\left(\frac{B_{i}\mu_{\mathrm{B}}+2I_{3i}\mu_{\mathrm{iso}}}{T}\right),
\eea
where $\mu_{\mathrm{B}}(\equiv 3\mu)$ and $\mu_{\mathrm{iso}}$ are 
the baryon and isospin chemical potentials, respectively.
Here we consider only the case of $\mu_{\mathrm{iso}}=0$.
The baryon number density is easily obtained as 
\bea
n_{\mathrm{B}}^{\mathrm{HRG}}=-\frac{\partial}{\partial \mu_{\mathrm{B}}}p^{\mathrm{HRG}}.
\eea
\\
\indent
There are lattice artifacts in LQCD simulations.
These were already discussed 
in Refs.~\cite{Huovinen,Philipsen,Allton,Ejiri,Bazavov}. 
For small $N_{t}$, for example, thermodynamic quantities exceed 
the Stefan-Boltzmann (SB) limit. 
Since it is not easy to eliminate the lattice artifact exactly, 
we take the following simple prescription. 
We consider the lattice SB limit that is defined by the lattice action 
with massless and free quarks, and 
normalize LQCD results with the corresponding 
values in the lattice SB limit in order 
to reduce the lattice artifacts; see Appendix~\ref{Quark number density in the free gas limit for the Wilson fermion} 
for the quark number density in the lattice SB limit. 
For the HRG model, meanwhile, the quark number density is normalized by 
the value in the continuum SB limit. 
In the HRG model, we assume that nucleon mass $m_{N}$ 
and $\Delta$-resonance mass $m_{\Delta}$ depend on $T$ only.
The baryon masses are determined so as to reproduce the LQCD result. 
Here we assume that the masses of 24 resonance states 
above the mass threshold $m_{cut}^{B}$ 
are fixed at 1.8 GeV, following Ref.~\cite{Huovinen}. 
But the contribution of 24 states to $n_q$ is small. 
\\
\indent
Figure~\ref{HRG_im-nmb} shows $\theta$ dependence 
of the normalized quark number density $n_{q}/n_{\mathrm{SB}}$ 
for $T=0.93T_{\mathrm{c}0}$ and $0.99T_{\mathrm{c}0}$.
The HRG-model results (solid lines) well reproduce 
$\theta$ dependence of LQCD results (symbols with error bars) 
for both cases of $T=0.93T_{\mathrm{c}0}$ and $0.99T_{\mathrm{c}0}$.
This implies that $m_{N}$ and $m_{\Delta}$ little depend on $\theta$. 
The resulting nucleon and $\Delta$-resonance masses  
are shown in Table~\ref{nucl-delta_mass}, together with $\chi^{2}/$d.o.f. 
values. The $\chi^{2}/$d.o.f. values are close to those for $g_{F}^2$ 
in Table \ref{Fourier_coeff}. 
The resulting masses are heavier than the corresponding physical values, 
because the quark mass is much heavier than the physical value in our simulations. As shown in Table~\ref{nucl-delta_mass}, 
both $m_{N}$ and $m_{\Delta}$ decrease by about 10\% 
as $T$ increases from $0.93T_{\mathrm{c}0}$ to $0.99T_{\mathrm{c}0}$.

\begin{table}[h]
\begin{center}
\begin{tabular}{cccc}
\hline
$T/T_{\mathrm{c}0}$ & $m_{N}$ & $m_{\Delta}$ & $\chi^{2}/$ d.o.f.
\\
\hline
\hline
~~$0.93$~~ & ~~$1091$~MeV~~ & ~~$1547$~MeV~~ & ~~$6.625$~~\\
~~$0.99$~~ & ~~~$940$~MeV~~ & ~~$1385$~MeV~~ & ~~$7.993$~~\\
\hline
\end{tabular}
\caption{
Results of $\chi^2$ fitting for baryon masses $m_{N}$ and $m_{\Delta}$ 
in the HRG model and $\chi^{2}/$d.o.f. values. 
}
\end{center}
\label{nucl-delta_mass}
\end{table}

\begin{figure}[h]
\begin{center}
\vspace{10pt}
\hspace{10pt}
 \includegraphics[angle=-90,width=0.445\textwidth]{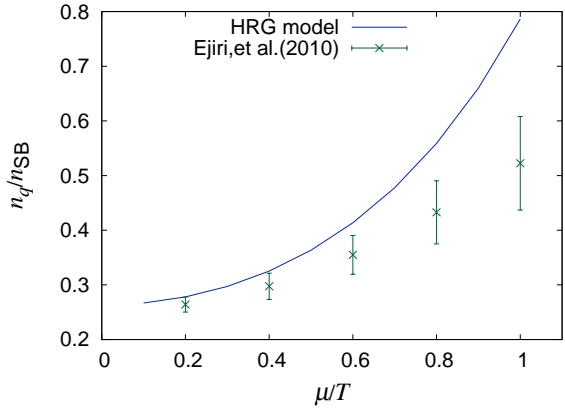}
\end{center}
\caption{
$\mu/T$ dependence of $n_{q}/n_{\mathrm{SB}}$ in the real $\mu$ region 
for $T=0.99T_{\mathrm{c}0}$. 
The symbols with error bars stand for LQCD results of 
the Taylor expansion method for the reweighting factor in Ref.~\cite{Ejiri}, 
while the solid line is the result of the HRG model. 
}
\label{HRG_re-nmb}
\end{figure}

Finally we consider the case of real $\mu$. 
The HRG model becomes less reliable as $\mu$ increases, because 
the pseudocritical temperature $T_{\mathrm{c}}(\mu)$ 
of deconfinement transition goes down from $T_{\mathrm{c}0}$ as $\mu$ increases from zero. 
Figure~\ref{HRG_re-nmb} shows the $\mu/T$ dependence of $n_{q}/n_{\mathrm{SB}}$ at $T=0.99T_{\mathrm{c}0}$. 
In the range $\mu/T < 0.4$, 
the HRG model result (solid line) is consistent with the previous LQCD 
results~\cite{Ejiri} (symbols with error bars)
based on the Taylor expansion method for the reweighting factor. 
Beyond the range, the HRG model overestimates the LQCD results. 
The HRG model is thus reliable only at $\mu/T < 0.4$ for the case of $T=0.99T_{\mathrm{c}0}$.

\section{Summary}
\label{Summary}
We have investigated $\mu$ dependence of $n_q$ at imaginary and real $\mu$, 
performing LQCD simulations at imaginary $\mu$ and extrapolating the results 
to the real-$\mu$ region by assuming functional forms for $n_q$. 
LQCD calculations were done on an $8^{2}\times 16\times 4$ lattice with the clover-improved two-flavor Wilson fermion action and the renormalization-group-improved Iwasaki gauge action. 
We considered two temperatures below $T_{\mathrm{c}0}$ and four temperatures above $T_{\mathrm{c}0}$. 
The quark number density was computed along the line of constant physics at $m_{\mathrm{PS}}/m_{\mathrm{V}}=0.80$.
\\
\indent
For imaginary $\mu$, the quark number density calculated with 
the Wilson-type fermion action is consistent with the previous 
result~\cite{D'Elia4} based on the staggered-type fermion action. 
The LQCD results thus do not depend on the fermion action taken. 
\\
\indent
We have extrapolated $n_q$ at imaginary $\mu$ to real $\mu$, 
assuming the Fourier series $g^{n}_{F}$ for $T<T_{\mathrm{c}0}$ and 
the polynomial series $g^{2n-1}_{p}$ for $T>T_{\mathrm{RW}}$; 
here the superscript $n$ represents the highest order in the partial sum. 
As for $T=0.99T_{\mathrm{c}0}$, the present result of $g^{2}_{F}$ is 
consistent with the previous result~\cite{Ejiri} of the Taylor expansion method for the reweighting factor in the range $\mu/T < 0.8$. 
The extrapolation based on $g^{2}_{F}$ is thus reliable at $\mu/T < 0.8$ 
for $T=0.99T_{\mathrm{c}0}$. 
Furthermore, 
the difference between the two results of 
$g^{1}_{F}$ and $g^{2}_{F}$ reduces as $T$ decreases from $T_{\mathrm{c}0}$, 
indicating that higher-order contributions become less important 
as $T$ decreases. 
Therefore, the extrapolation based on $g^{2}_{F}$ is reliable 
at $\mu/T < 0.8$ for any $T$ less than $T_{\mathrm{c}0}$.
\\
\indent
For $T>T_{\mathrm{RW}}$, 
the previous study based on the Taylor expansion method for the reweighting factor has contributions 
up to $(\mu/T)^3$, but the present $g^{5}_{p}$ extrapolation retains contributions up to $(\mu/T)^5$. 
Using this advantage of the present method from 
the previous method, we have estimated to what extent 
the Taylor expansion or the $g^{3}_{p}$ extrapolation works. 
The upper bound $(\mu/T)_{\rm max}$ of the reliable extrapolation 
goes up as $T$ increases from $T_{\mathrm{RW}}$, 
because higher-order contributions become less important. 
\\
\indent
The HRG model is reliable in the confinement region. 
For the 2+1 flavor case at zero chemical potential, in fact, 
the HRG model well reproduces LQCD data on 
pressure at $T<1.2T_{\mathrm{c}0}~$\cite{Borsanyi:2012cr}. 
When $\mu$ is varied with $T$ fixed at $T_{\mathrm{c}0}$, 
the system is in the confinement phase at imaginary $\mu$ but in 
the deconfinement phase at real $\mu$. 
This means that the HRG model is more reliable at imaginary $\mu$ than at 
real $\mu$, when $T$ is fixed at $T_{\mathrm{c}0}$. 
We have then tested whether $T$ dependence of $m_{N}$ and $m_{\Delta}$ 
in the vicinity of $T_{\mathrm{c}0}$ can be determined 
from LQCD data on $n_q$ at imaginary $\mu$ by using 
the HRG model. The HRG model well reproduces the LQCD results, 
when $m_{N}$ and $m_{\Delta}$ are assumed to depend on $T$ only. 
This implies 
that $m_{N}$ and $m_{\Delta}$ little depend on $\theta (=\mu_{\rm I}/T)$. 
In our test calculation, $m_{N}$ and $m_{\Delta}$ reduce by 
about 10\% when $T$ increases from 0.93$T_{\mathrm{c}0}$ 
to 0.99$T_{\mathrm{c}0}$. We propose this method as a handy way 
of determining $T$ dependence of $m_{N}$ and $m_{\Delta}$ 
in the vicinity of $T_{\mathrm{c}0}$. 
This method is practical, since it is not easy to measure $T$ dependence of 
pole masses directly with LQCD simulations.

\noindent
\begin{acknowledgments}
We thank A. Nakamura and K. Nagata for useful discussions and giving the LQCD program codes.
We are also grateful to I.-O. Stamatescu, T. Sasaki and T. Saito for helpful discussions. 
J. T., H. K., and M. Y. are supported by Grant-in-Aid for Scientific Research (No. 25-3944, No. 26400279 and No. 26400278) from the Japan Society for the Promotion of Science (JSPS).
The numerical calculations were performed on NEC SX-9 and SX-8R at CMC, Osaka University and on HITACHI HA8000 at Research Institute for Information Technology, Kyushu University.
\end{acknowledgments}

\noindent
\appendix
\section{Quark number density for the Wilson fermion in the massless free-gas limit }
\label{Quark number density in the free gas limit for the Wilson fermion}
We consider $n_q$ for the Wilson fermion 
in the lattice SB limit (the massless free-gas limit). 
In the high-$T$ limit, we can consider a quark as a massless and 
non-interacting particle,  
since the effects of finite quark mass and interactions between quarks are 
negligible there. 
In Appendix of Ref.~\cite{Ejiri}, the lattice SB limit is discussed except for 
the quark number density.

The partition function with free Wilson fermions is given by
\bea
Z(\kappa,\hat{\mu}) &=& (\det M )^{N_f}, \\
M_{xy} &=& \delta_{xy} - \kappa \sum_{i=1}^{3}\left[ (1-\gamma_{i})\delta_{x+\hat{i},y}+ (1+\gamma_{i})\delta_{x-\hat{i},y} \right] \nonumber \\
&& - \kappa \left[ e^{+\hat{\mu}}(1-\gamma_{4})\delta_{x+\hat{4},y}+ e^{-\hat{\mu}}(1+\gamma_{4})\delta_{x-\hat{4},y} \right], \nonumber \\
&&
\eea
on an $N_{x}\times N_{y}\times N_{z}\times N_{t}$ lattice. 
After the unitary transformation to momentum space, we obtain
\bea
Z(1/8,\hat{\mu}) &=& \left( \prod_{k} \det \tilde{M}(k) \right)^{N_{c}N_{f}} \\
\det \tilde{M}(k) &=& \frac{16}{8^{4}} \left[ \sum_{i} \sin^{2}k_{i} + \left\{ 2\sum_{i}\sin^{2} \left( \frac{k_{i}}{2} \right) \right\}^{2} \right. \nonumber \\
&& \left. + 4\left\{ 2\sum_{i}\sin^{2} \left( \frac{k_{i}}{2} \right) + 1 \right\} \sin^{2}\left( \frac{k_{t}-i\hat{\mu}}{2} \right) \right]^{2} \nonumber \\
\label{eq:12}
\eea
in the massless quark limit $\kappa=1/8$, where $\tilde{M}$ is 
the fermion matrix in momentum space, and 
\bea
 k_{i} = \frac{2\pi j_{i}}{N_{i}}, \; j_{i} = 0, \pm 1, \cdots, N_{i}/2
\eea
for the spatial components ($i=x,y,z$) and
\bea
 k_{t} = \frac{2\pi(j_{t}+1/2)}{N_{t}}, \; j_{t} = 0, \pm 1, \cdots, 
 N_{t}/2
\eea
for the time component. 
The quark number density in the lattice SB limit is then obtained as 
\bea
\frac{n_{q}}{T^{3}} &=& \frac{N_{t}^{3}}{N_{V}} \frac{\partial}{\partial \hat{\mu}}\ln Z(1/8,\hat{\mu}) \nonumber \\
&=& N_{c}N_{f}\frac{N_{t}^{3}}{N_{V}} \sum_{k} \frac{\partial \det \tilde{M}(k)}{\partial \hat{\mu}} \left[ \det \tilde{M}(k)\right]^{-1} ~~~ 
\label{eq:15}
\eea
with 
\bea
\frac{\partial \det \tilde{M}(k)}{\partial \hat{\mu}} &=& -\frac{1}{8^{2}}\left\{ 2\sum_{i}\sin^{2} \left( \frac{k_{i}}{2} \right) + 1 \right\} \nonumber \\
&& \times (\sinh \hat{\mu} \cos k_{t} + i \cosh \hat{\mu} \sin k_{t} ).~~~
\label{eq:16}
\eea
The quark-number density at imaginary $\mu$ is obtained by replacing 
$\hat{\mu}$ by $i\hat{\mu}_{\mathrm{I}}$ in Eqs. \eqref{eq:15} and 
\eqref{eq:16}.


\end{document}